\documentclass[%
 aip,
 apl,%
 amsmath,amssymb,
 preprint,groupedaddress
]{revtex4-1}

\usepackage{graphicx}
\usepackage{dcolumn}
\usepackage{bm}
\usepackage{amssymb,multirow,wrapfig}
\usepackage{color}
\usepackage{graphicx}

\newcommand{\abs}[1]{\ensuremath{\left| #1 \right|}}

\newcommand{\ket}[1]{\ensuremath{\left| #1 \right>}}

\newcommand{\be}[0]{\begin{equation}}
\newcommand{\ee}[0]{\end{equation}}
\newcommand{\bea}[0]{\begin{eqnarray}}
\newcommand{\eea}[0]{\end{eqnarray}}

\newcommand{\sbsc}[1]{\ensuremath{_{\textrm{#1}}}}

\begin{document}


\title{Quantum-Secured Imaging} 



\author{Mehul Malik}
\email[]{memalik@optics.rochester.edu}

\author{Omar S. Maga\~{n}a-Loaiza}

\author{Robert W. Boyd}
\altaffiliation[Also at ]{Department of Physics, University of Ottawa, Ottawa, ON K1N 6N5 Canada}
\affiliation{The Institute of Optics, University of Rochester, Rochester, NY 14627}


\date{\today}

\begin{abstract}
We have built an imaging system that uses a photon's position or time-of-flight information to image an object, while using the photon's polarization for security. This ability allows us to obtain an image which is secure against an attack in which the object being imaged intercepts and resends the imaging photons with modified information. Popularly known as ``jamming," this type of attack is commonly directed at active imaging systems such as radar. In order to jam our imaging system, the object must disturb the delicate quantum state of the imaging photons, thus introducing statistical errors that reveal its activity. 
\end{abstract}

\pacs{}

\maketitle 
Recent advances in quantum mechanics have enabled many enhanced imaging technologies \cite{Kolobov:1999ex,Lugiato:2002wa}. Entangled photon-number ($N00N$) states \cite{Boto:2000eg} have allowed Heisenberg-limited phase measurement and led to the development of LIDAR systems with quantum-enhanced resolution \cite{Giovannetti:2004jg}. Even without the use of entanglement, the sensitivity of optical ranging and pointing systems has been improved beyond the classical limit by the use of quantum resources \cite{Spagnolo:2012fc,Dutton:2010fj, Treps:2003ho}. In this letter, we propose and demonstrate a quantum enhancement to optical ranging and imaging systems that will make them secure against intercept-resend jamming attacks. A common concern for active imaging systems today is the threat of jamming, where extraneous or false information is sent to the receiver in order to fool it \cite{EWInfo:1999}. More sophisticated methods of jamming are being developed which allow the imaging signal to be intercepted, manipulated, and resent \cite{Roome:1990th}. This allows the object being imaged to bely its actual position or velocity, or even create a false target \cite{Kristoffersen:2004uu}. By using quantum states of light modulated in polarization in an imaging system, we can provide security against such methods of jamming. Quantum-secured sensing based on similar principles has previously been demonstrated for the purpose of sensing intruders by using entanglement \cite{Humble:2009dc} and interaction-free measurements \cite{Anisimov:2010cn}.

\begin{figure}[b!]
\centering\includegraphics[scale=.7]{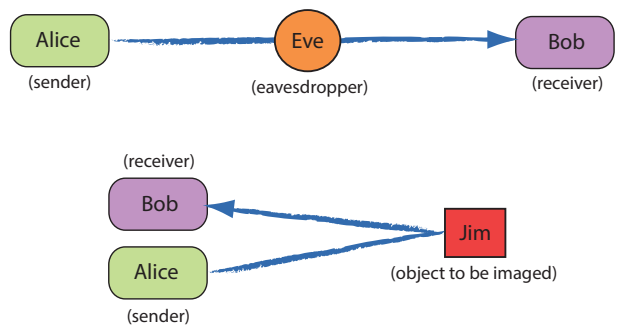}
  \caption{A sketch showing the fundamental difference between the quantum key distribution (QKD) and quantum-secured imaging (QSI) protocols. In QKD, a spatially separated sender and receiver use quantum mechanical principles to securely share information. In QSI, a collocated sender and receiver use shared information to securely query an object.}\label{SQIvQKD}
\end{figure}

Our secure imaging technique is based on a modified version of the BB84 protocol of quantum key distribution (QKD) \cite{Bennett:1984wv}. Instead of an eavesdropper (Eve) located between the sender (Alice) and the receiver (Bob), we now have a jamming object (Jim) at one end and Alice and Bob at the other (Fig. \ref{SQIvQKD}). By virtue of being in the same location, Alice and Bob already share information. Instead, they now use this information to securely query Jim by encoding it in the polarization of a stream of photons. This leaves the position and time degrees of freedom of the photons free for the purpose of obtaining an image of Jim. If Jim were to try to jam this system by intercepting and resending the photons with false position or time information, he will introduce statistical errors in the polarization encoding that will give away his jamming attempt. As in QKD, security is guaranteed due to Jim's inability to measure a photon simultaneously in two conjugate polarization bases.

Studies of eavesdropping in QKD  \cite{Huttner:uz,Bourennane:2002uo} attempt to answer the question: what is the maximum error rate detected by Bob that will allow the extraction of secure information after error correction and privacy amplification? To jam our secure imaging protocol, the jamming object, Jim, must perfectly replicate our entire querying signal in order to resend it with false position or time information. This simplifies the above question to: what is the minimum error rate introduced by Jim in trying to copy a secure QKD transmission between Alice and Bob? Using the intercept-resend quantum eavesdropping strategy \cite{Bennett:1992ws,Gisin:2002zz}, Jim can pick two orthogonal polarization bases to eavesdrop in. His error rate is then equal to:

\be e_{\textrm{J}}(\theta)=\frac{1}{4}\big[(1-\cos 2\theta)+(1-\sin 2\theta)\big] \ee

\noindent where $\theta$ is the angle between the preparation basis used by Alice and the eavesdropping basis. Jim's error rate ($e_\textrm{J}$) is minimized to $14.64$\% when $\theta=22.5^{\circ}$ (referred to as the Breidbart basis \cite{Bennett:1982tq}). However, Bob's error rate ($e_\textrm{B}$) is independent of the jamming basis angle used and is minimized to 25\% as long as Jim always resends in the eavesdropping basis \cite{Bennett:1992ws}: 

\be e_{\textrm{B}}(\theta)=\frac{1}{4}\big[(1-\cos^2 2\theta)+(1-\sin^2 2\theta)\big] = 25\%.\ee

\noindent We use this error rate as our secure image error bound. If Bob's received signal has an error rate less than 25\%, images obtained from it can be considered secure against intercept-resend jamming attacks. Images obtained from a signal with an error rate greater than 25\% cannot be considered secure and imply that Jim was actively jamming the channel. This can be interpreted as a reduction in Alice and Bob's mutual information, which is related to Bob's error rate as:

\begin{figure}[t!]
\centering\includegraphics[scale=.68]{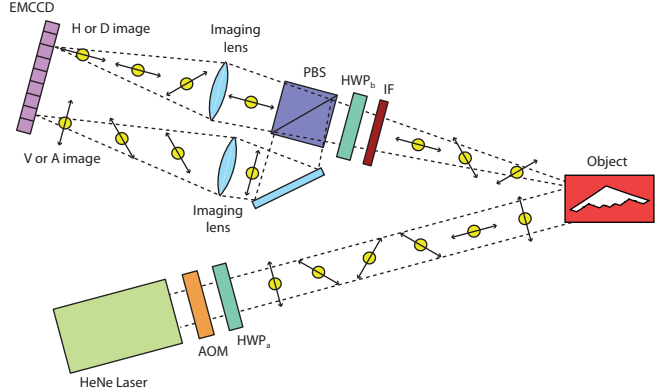}
  \caption{Schematic of our quantum-secured imaging experiment. Polarized single-photon pulses from a HeNe laser are reflected from the object and imaged onto an electron-multiplying camera (EMCCD) through an interference filter (IF). A half-wave plate (HWP) and a polarizing beamsplitter (PBS) are used to make the appropriate polarization basis measurement. Four images corresponding to the four measured polarizations are obtained. The angle of reflection is exaggerated in the figure for clarity but is less than 5$^{\circ}$ in reality. The object consists of a reflective stealth aircraft silhouette.}\label{setup}
\end{figure}

\be I_{\textrm{AB}}=1+(1-e_{\textrm{B}})\log_2(1-e_{\textrm{B}})+e_{\textrm{B}}\textrm{log}_2(e_{\textrm{B}}).\label{Iab}\ee

\noindent For our imaging protocol to be secure, Alice and Bob's mutual information after querying Jim must be at least 0.1887 bit/photon. In a protocol with no error, their mutual information stays at its maximum value of 1 bit/photon.

In our proof-of-principle experiment, we use polarization-modulated photons to securely image an object in reflection (Fig. \ref{setup}). A HeNe laser is intensity modulated by an acousto-optic modulator (AOM) to create pulses with one detected photon on average. A half-wave plate (HWP$_\textrm{a}$) mounted on a motorized rotation stage randomly switches the polarization state of the photon among horizontal, vertical, diagonal, and anti-diagonal (\ket{H}, \ket{V}, \ket{D}, and \ket{A}). The single-photon pulses are incident on the object, which consists of a stealth aircraft silhouette on a mirror. They are then specularly reflected from the object towards our detection system. In Fig. \ref{setup}, we show a non-zero reflectance angle for clarity. An interference filter (IF) is used to eliminate the background. A second rotating half-wave plate (HWP$_\textrm{b}$) and a polarizing beam-splitter (PBS\sbsc{b}) carry out the appropriate basis measurement. When the axis of HWP$_\textrm{b}$ is parallel or at $22.5^{\circ}$ to the H direction, the measurement is carried out in the horizontal-vertical ($H/V$) or diagonal-anti-diagonal ($D/A$) basis respectively. If an \ket{H} or \ket{V} photon is sent, the measurement on the received photon is always carried out in the $H/V$ basis, and similarly for the $D/A$ basis. This removes the need for sifting bases between sent and received photons. Two lenses are used after the PBS to create four images corresponding to the four measured polarizations on an electron-multiplying CCD camera (EMCCD), which serves as a single-photon detector.

\begin{figure*}
\centering\includegraphics[scale=.78]{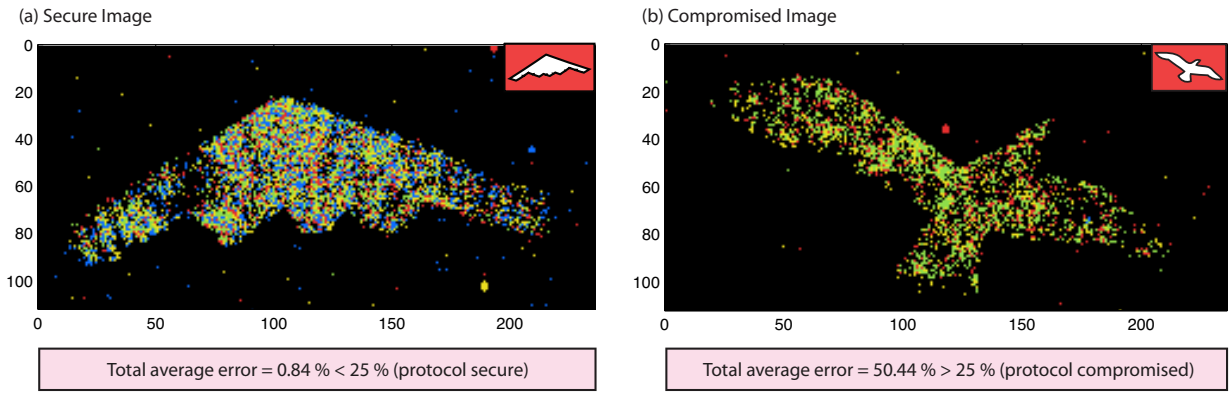}
  \caption{Laboratory demonstration of quantum-secured imaging. (a) When there is no jamming attack, the received image faithfully reproduces the actual object, which is shown in the inset. (b) In the presence of an intercept-resend jamming attack, the received image is the ``spoof" image of a bird. However, the imaging system can always detect the presence of the jamming attack, because of the large error rate in the received polarization. In (a) the error rate is 0.84\%, while in (b) it is 50.44\%. A detected error rate of $>25\%$ indicates that the image received has been compromised.}\label{image_data}
\end{figure*}

\begin{figure}[t!]
\centering\includegraphics[scale=.78]{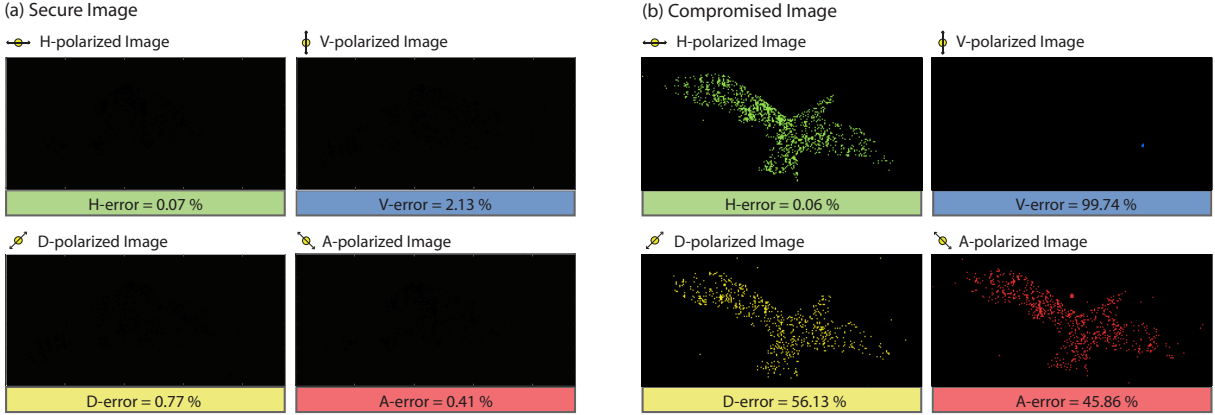}
  \caption{The received image is comprised of four different images corresponding to the four measured polarizations ($H, V, D$, and $A$). (a) When there is no jamming attack, the four images have a near-zero error in the received polarization. (b) In the presence of an intercept-resend jamming attack in which the object resends only $H$-polarized photons, the four images have considerable error in the received polarization. This measured error allow Alice and Bob to determine that the imaging system was being actively jammed. There is no $V$ image obtained in this case as the measurement of an $\ket{H}$ photon in the $HV$ basis leads to no $V$ signal.}\label{pol_data}
\end{figure}

In Fig. \ref{image_data}(a), we show an image of the stealth aircraft object obtained by this system. The image is obtained by taking 10,000 images containing one detected photon each on average, for random orientations of HWP\sbsc{a} and HWP\sbsc{b}. The final image is constructed by combining the four polarization images formed on the EMCCD, shown in Fig. \ref{pol_data}(a). The different pixel colors correspond to the different measured polarizations. An error corresponds to the case when a received photon is detected in the opposite polarization to that it was sent in. For example, if an \ket{H} photon is sent to the object and a click is obtained in the \ket{V} image, it counts as an error. For the case when there is no jamming attack, we expect an error-free image. However, some error is obtained due to imperfections in the PBS, and is in agreement with the measured PBS efficiencies. The measured average error of 0.84\% is well below our error bound of 25\%, indicating a secure image. Using Eq. \ref{Iab}, we see that the imaging system's mutual information is reduced to 0.93 bit/photon, which is above the threshold value of 0.1887 bit/photon.

We simulate an intercept-resend jamming attack on our system by intercepting the imaging photons at the object, and resending them with a ``spoof" image of a bird. For simplicity, we resend all the photons in a horizontal polarization. Fig. \ref{image_data}(b) shows the received image in this case. The presence of the jamming attack is detected by measuring the error in each received polarization. Measurements of an \ket{H} photon in the $D/A$ basis give an average error of 50\%. Measurements of an \ket{H} photon in the $H/V$ basis always appear in the $H$ channel. Thus, when a \ket{V} photon is expected, there is a 100\% error. When an \ket{H} photon is expected, no error is obtained. These error probabilities give an average expected error of 50\%. Our measured error of 50.44\% closely matches this result and indicates that the received image has been compromised. Also, the system's mutual information is reduced to near-zero, further verifying an intercept-resend jamming attack. We show the four polarization images and their measured errors in polarization in Fig. \ref{pol_data}(b). 

\begin{figure}[t!]
\centering\includegraphics[scale=.68]{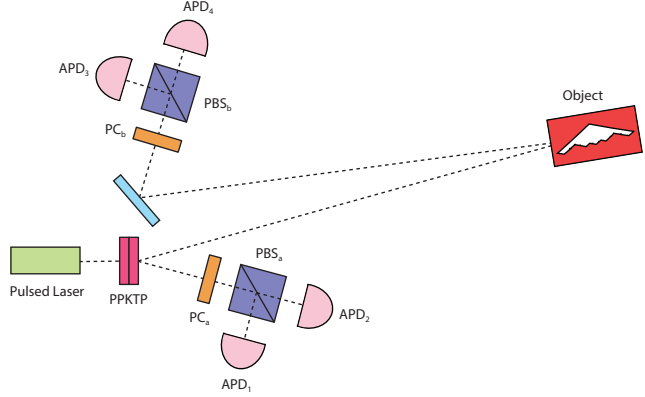}
  \caption{Schematic for a proposed secure time-of-flight experiment, based on the entanglement-based Ekert QKD protocol. Polarization-entangled photon pairs generated in a pair of crossed PPKTP crystals are used to measure the distance to an object. Security against an intercept-resend jamming attack is checked by carrying out a test for a CHSH Bell inequality with measurements in appropriate polarization bases using Pockels cells (PC) and polarizing beam-splitters (PBS).}\label{lidar}
\end{figure}

While we have performed secure imaging using a photon's position information, it is easy to extend this idea to a photon's time-of-flight information. In addition, one can use the entanglement-based Ekert protocol for security. In Fig. \ref{lidar} we propose a schematic for an entanglement-based secure optical ranging experiment. A pulsed laser incident on a pair of crossed periodically poled potassium titanyl phosphate crystals (PPKTP) creates pulses with one pair of polarization-entangled photons on average, in the state $(\ket{H_1H_2}+\ket{V_1V_2})/\sqrt{2}$. Using an appropriately oriented Pockels cell (PC), a PBS, and two avalanche photodiodes (APDs), these photons are measured in the rotated polarization basis $\ket{H'}+\ket{V'}$, where $\ket{H'}=\sin\theta\ket{V}+\cos\theta\ket{H}$ and $\ket{V'}=\cos\theta\ket{V}-\sin\theta\ket{H}$. One photon from the polarization-entangled pair is immediately measured by PC\sbsc{a}, PBS\sbsc{a}, APD\sbsc{1}, and APD\sbsc{2} in one of two rotated polarization bases with $\theta = 0^{\circ}$ and $45^{\circ}$. The other photon travels to the object and is reflected back to the source, where it is measured by PC\sbsc{b}, PBS\sbsc{b}, APD\sbsc{3}, and APD\sbsc{4} in one of two rotated polarization bases with $\theta = 22.5^{\circ}$ and $-22.5^{\circ}$. For each pulse, coincidence timing measurements between APDs\sbsc{1,2} and APDs\sbsc{3,4} are used to calculate the CHSH Bell inequality parameter $S$ \cite{Kwiat:1995ub,Jha:2008wu}, as well as the distance to, or velocity of the object. If the calculated CHSH parameter meets the condition $\abs{S}>2$, the optical ranging measurement can be considered secure against an intercept-resend jamming attack. Such a technique would greatly enhance the security of photon-counting optical ranging systems being developed today \cite{McCarthy:2009te,Howland:2011wo}.

In conclusion, we have implemented an active imaging scheme that uses quantum mechanical principles to ensure security against intercept-resend jamming attacks. We have also proposed a quantum-secured optical ranging technique. We should point out that our proposed schemes have certain limitations. Our experimental implementation used weak coherent pulses, which makes it susceptible to a photon-number splitting attack, where the jammer splits one or more photons from pulses containing more than one photon \cite{Brassard:2000io}. This would allow the jammer to measure these photons in both polarization bases and perfectly replicate the querying pulses. It may be possible to use decoy states to defeat such an attack, as has been demonstrated in QKD \cite{Zhao:2006dc}. Further, a sophisticated jammer may use quantum teleportation \cite{Bouwmeester:1997wk} to teleport the polarization state of our querying photons onto photons carrying false position or time information. In practice, however, this would prove extremely challenging, as quantum teleportation involves Bell state measurements, which can only be performed probabilistically in a linear optical scheme \cite{Humble:2009dc}. Finally, our protocol does not provide security against attacks that preserve a photon's polarization state. For example, metamaterials \cite{Pendry:2006hq} and slow-light techniques \cite{Bigelow:2003wr} can be used to hide an object in space and time without disturbing the polarization of any querying photons. However, these methods are currently in their infancy and only work in extremely limiting cases. On the other hand, given the current state of QKD technology \cite{Dixon:2010ev, Steinlechner:2012ex}, our quantum-secured protocol can easily be realized and integrated into modern optical ranging and imaging systems. Also, the possibility of using other degrees of freedom of a photon such as its orbital angular momentum in a quantum-secured channel \cite{Malik:2012ka} may open up exciting avenues for future research.

\begin{acknowledgments}
We would like to thank Dr. J. Kuper, Dr. J. Leach, Dr. J. C. Howell, and A. C. Liapis for helpful discussions. This work was supported by the DARPA InPho program and the CONACYT.
\end{acknowledgments}

%

\end{document}